# Consider ethical and social challenges in smart grid research

Valentin Robu, David Flynn, Merlinda Andoni, Maizura Mokhtar

Heriot-Watt University, Edinburgh, Scotland, UK

Contact email: v.robu@hw.ac.uk

*Artificial Intelligence and Machine Learning are increasingly seen as key technologies for building more decentralised and resilient energy grids, but researchers must consider the ethical and social implications of their use*

Energy grids are undergoing rapid changes, requiring new ways both to process the large amounts of data generated from the power system, but also – increasingly – to take smart operational decisions [1].

On the data side, the UK and most EU countries have committed to a target of offering a smart meter to every home by 2020 [2], with similar monitoring being installed in other parts of the energy network. This has led to some to refer to a "data tsunami", requiring development of new machine learning techniques to deal with the ensuing challenge of extracting useful information from this data – often in real time.

Another trend is the use of AI techniques (such as those from multi-agent systems, computational game theory and decision making under uncertainty) to take autonomous allocation and control decisions. This is driven increasingly by the moves towards more decentralised energy systems, where prosumers (consumers with own micro-generation and storage) can generate and source their own electricity through peer-to-peer (P2P) trading in local energy markets and community energy schemes.

To illustrate these trends from a UK perspective (with which the authors are most familiar), a number of projects are looking at computational techniques suitable for more decentralised energy future – including CESI (the UK National Centre for Energy Systems Integration) [23] and Responsive Flexibility, one of the largest smart energy demonstrators in the UK, focused on the islands of Orkney [26]. In fact, Community Energy Scotland (an organisation supporting such projects in Scotland) lists no less than 300 projects on their website [3]. This has attracted a surge of interest and considerable investments in technologies enabling these settings, such as blockchain, machine learning (ML) and distributed Artificial Intelligence (AI) [4].

Yet, there are substantial ethical and social questions to be asked in such developments. The AI community has begun to recognise the challenges posed by rapid adoption of AI in the real world, and a number of general guidelines have been proposed [5]. Similarly, the smart grid community needs to consider the ethical challenges that the rapid adoption of AI and ML techniques to control our energy systems brings. In this piece, we discuss some domain-specific challenges and proposed solutions for adopting AI techniques to automate smart energy grids. We note our starting point is not purely generic AI ethical principles (although we have consulted recent reports, such as Floridi [5], which provide a useful guide). Rather, we consider the trends that have shaped AI adoption in smart grids in recent years, and illustrate the type of ethical challenges involved in each.

**Privacy-preserving analysis of energy data**

One active area of research effort (in particular from the ML community) in recent years has been the development of new algorithms for energy disaggregation or appliance recognition [6,7]. Essentially, the aim of such algorithms is to use fine-grained data from a smart meter on the aggregate energy consumption of a household, to deconstruct and identify what appliances are being used at any one time. This can be very useful for several applications, including billing, micro-energy trading, demand-side response, but it also raises a number of privacy concerns.

In fact, in the UK, the national regulator (OFGEM) asks utilities and network operators to sign privacy pledges, to guarantee the security and privacy of their customers. We argue that, in the future, such regulation will need to be increasingly informed by the ability of the latest AI/ML techniques. Recent research funded by network operators (such as Scottish Power Networks) [8] is increasingly looking at how ML techniques such as deep learning, can be used to leverage smart meter data to address power network challenges (for example, voltage exceeding safe limits), but in a privacy-preserving way, that only need to collect data from a few key identified locations, rather than all users in the network.

**Incentive design in energy systems**

One very active area for recent smart grid research has been in optimisation and scheduling of microgrids and local energy systems. Such applications usually consider a system composed of a number of agents, often with their own self-interest and different objectives. One the one side, we have the local consumers (which often have an EV) and which require some amount of electricity to run their jobs (e.g. EVs getting charged) – from their perspective, at the lowest possible price and earliest opportunity. On the other hand, there may be individual prosumers with small renewable generation units (e.g. micro wind turbines or solar panels) and/or battery storage, who aim to profitably trade their excess renewable generation or capacity. Finally, there are the system operators and managers of local energy marketplaces and community energy schemes. Their objectives are often twofold: first, they must assure the stability of the local power system, and second they want to optimise the system such that the final allocation is efficient and fair according to some criteria. Yet, in designing a distributed local energy system, they need to consider not just a global optimality, but the preferences and behaviour of individual agents, which might not be aligned with those of the overall system designer.

Consider a local community in which many households buy an electric vehicle (EV). If all of them charge at the same time, the constraints of the local distribution grid may be exceeded, so it would be natural to schedule first EVs with more urgent deadlines. Yet, such a mechanism needs to be complemented by incentive schemes that assure, for example, that EV owners who are willing to be patient are rewarded with cheaper electricity. The distributed AI community has a key role to play in using algorithmic game theory, to design automated markets in which intelligent agents control the EV charging for their owners [9].

Other relevant examples include demand-side response [10,11], where consumers are rewarded to temporarily reduce their demand. The design of these reward systems needs to consider both the optimisation of system performance, but also use multi-agent and game theory to design allocation mechanisms that balance cost savings with the risk that agents that commit to respond may not be able to do so.

**Blockchains and distributed AI**

One trend that has emerged very recently is the interest in distributed AI and blockchain technology. Our recent review supported by the UK National Centre for Energy Systems

Integration [4] identifies an explosion of interest in this space, with no less that 140 projects and start-ups looking at such settings, and a recent US Congressional Research Office [12] report highlights the importance of this technology in the energy space. While blockchains do have clear advantages in terms of assuring the traceability and enforcing transactions without requiring an intermediary like a utility company, they need to be complemented by AI/multi-agent solutions to realise their full potential. Specifically, many blockchain solutions work by allowing smart contracts to be used to enable peer-to-peer trading in distributed microgrids and energy communities. Yet, designing and agreeing these smart contracts requires tools from sub-fields of AI, such as automated, agent-mediated negotiation and algorithmic mechanism design [9-11]. Moreover, it often requires the AI-enabled system to make judgments on, for example, which prosumers get priority in power allocation in constrained networks and subsequently get paid for their micro-renewable generation, hence inputs from social sciences are highly valuable.

**Social science perspectives**

The importance of considering social science input when designing applications for AI systems has been noted by a number of authors, e.g. [13]. This is of great importance in smart grid applications as well. A key reason for this is that, even if the latest multi-agent systems and game-theoretic tools are used to automate such settings, people's energy practices may differ significantly from game-theoretic models. Moreover, the benefits of the technology, for example the benefits that AI-enabled smart contracts and local energy markets offer may not be understood by all users, which can leave some users behind or create resentments when smart grid solutions are deployed on the ground.

Recent efforts from the AI and Human Computer Interaction communities have so far focused in two aspects. First, in designing novel ways to interact with energy, such as intuitive interfaces and devices that enable consumers to specify their energy preferences, understand their own energy consumption patterns [14,15], and automatically search for tariffs that best fit their energy needs [16]. Second, social scientists have begun to look at novel ways to understand consumers' energy practices. Such research is often more qualitative than the quantitative type of research done in AI, but can provide valuable insights for AI system design.

**Fair algorithmic decision-making**

Overall, AI in energy systems is being used to take increasingly complex decisions. For example, when designing a smart contract in a local energy market, this contract will specify who gets access to energy first (e.g. which EV owner can charge first and who needs to wait [9]), or conversely, in case there is excess renewable energy, which wind turbines should get priority to sell their energy or might need to curtail their generation (e.g. [17]). Using AI systems to take decisions clearly involves not just economics, but also social and ethical considerations in the way systems are designed and programmed, hence the importance of inputs from these disciplines.

**Developing country grids**

An interesting aspect is that AI techniques have not been used just for automating the energy system of developed countries, but a significant body of works is increasingly considering developing countries' grids. In such settings, which include many countries in SE Asia such as India or sub-Saharan Africa, where communities have to rely on local embedded micro-generation and storage for their energy needs. P2P energy trading mechanisms and local energy schemes are a natural fit. For example, a household with a large solar panel can often make additional income by selling excess generation to its neighbour without one in times of excess supply, and conversely, buy energy from another neighbour's battery in times of shortage.

The AI community has begun investigating some of these challenges. For example, in recent joint work with Covenant University of Nigeria, our group has investigated how automated agents and agent-mediated electronic negotiation (such as searching for Pareto-optimal deals with incomplete information) can be used to best automated P2P micro energy transactions [19], building on the results of a previous study by Alam et al. [18].

Recent work in the group of Professor Abhyankar at IIT Delhi [21] has proposed use of Shapley values (a coalitional game theory technique) for for loss allocation in meshed distribution networks, while Norbu & Bandyopadhyay [22], working at the Royal University of Bhutan, have investigated how pinch point analysis can be used to determine the optimal sizing of renewable generation in isolated distribution systems. Another developing-country specific application is considered in Ramchurn et al [20], who use AI-based optimisation to schedule load shedding (i.e. planned blackouts) in countries with weak grid environments.

In such applications, where machines take critical decisions about users' energy consumption, it is critical that input from social science and surveys on the acceptability of the system conducted with users impacted by the system are taken into account. This is the perspective taken in a number of energy projects, such as EPSRC CEDRI [24], a large-scale joint UK-India collaboration into smart energy systems.

**Robust and safe next-generation grids**

We conclude that many of our energy systems are becoming increasingly automated. For example, the UK government plans to supply a 1/3 of its electricity form offshore wind turbines [25], which are often monitored by autonomous robots and managed remotely by increasingly complex, AI-enabled software. Yet, as the recent blackout event in the UK has shown, careful control of such systems is needed, and the safety of these AI systems needs to be ensured. The AI community has a key role to play, not just in assuring that tomorrow's smarter grid systems run more efficiently, but also that they will be trustworthy and incorporate ethical and social principles in how they treat the preferences and constraints of their users (be they consumers, generators or network operators).

**Authors and affiliations:**

**Valentin Robu, David Flynn, Merlinda Andoni, Maizura Mokhtar**
Heriot-Watt University, Edinburgh, Scotland, UK
Contact email: v.robu@hw.ac.uk

**Acknowledgments:**
The authors would like to acknowledge the support of the UK Research Councils through the CESI projects CESI [EP/P001173/1], CEDRI [EP/R008655/1], NCEWS [KTP:510925] and Reflex [InnovateUK: 104780].

**Competing interests statement:**
The authors declare no competing interests.